# [1]Multi-Cell Traps for Registering of Cold Atom Clouds

G.A. Kouzaev and K.J. Sand

Department of Electronics and Telecommunications
Norwegian University of Science and Technology
O.S. Bragstads plass 2B, Trondheim, No-7491

## Abstract

In this letter, the results on the development and simulation of new three-dimensional nanotraps for cold dressed atoms are considered. The traps are the multi-cell structures built by crossed non-touching carbon nanotubes. The trapping effect is tuned by the DC and RF currents to confine the strong- or low-potential seeking atoms far enough from the areas of strong Casimir-Polder and spin-flip forces. It is supposed that the developed and simulated multi-cell structures are pertinent for catching the Bose-Einstein condensates to demonstrate the Josephson effect, and to enable the study of the entanglement of confined clouds in three-dimensional nano-cells.

## 1. Introduction

Trapping of cold atoms by static electric and magnetic (DC) fields is a promising technology to study the quantum nature of matter [1],[2]. In these traps cold alkali atoms interact with the fields due to their spins or/and electrical dipole moments. This interaction is described with the effective potential. The potential maxima are sought by the ground state atoms. Some sorts of excited atoms are attracted to the potential minima. Very often, the potential maxima are glued to the conductor surface, and the ground-state atom trapping cannot be realized.

More flexibility is provided by the dressing effect when the atoms interact with large quanta of the electromagnetic field [3]. Some such traps of this type are realized by combinations of strong DC and radio-frequency (RF) fields, and different theoretical models of them have been published recently [4]-[12]. For example, an analytical formula derived in [6] for the effective potential was verified by measurements for several traps and interferometers. In [7], a theory for a multi-minimum highly-versatile potential generated by DC and multi-frequency currents is given. The RF Ioffe-Pritchard trap [8]-[10] confines two sorts of atoms in contrast to the traps with only DC excitation. The weak-field-seeking atoms concentrate at the effective potential minima. The strong-field-seeking atoms stay at the potential maxima. This trap is prospective for the study of interactions of two sorts of atoms and the subsequently arising quantum entanglement. New potential shapes are studied in [11],[12] for two crossed wires carrying DC and RF currents. Each wire is surrounded by a cylindrically formed potential minimum that traps the cold atoms. These potential minima can touch each other under certain conditions for the DC and RF currents. If outer biasing field is varied slowly, the atoms can be pushed adiabatically from one wire to another. This studied configuration is interesting for the addressed transportation of atoms and for quantum interferometry.

The most developed and studied traps are on the scale of centimeters in size. Recent traps are based on micron-sized integrated technology, and they catch millions of alkali atoms [13]. New results shows miniaturization down to the nanoscale range taking into account several effects limiting proximity of cold atom clouds to the nanoconductors [14], [15]. Nanotubes can also be filled by a magnetic material or be magnetic themselves to decrease the overall trapping DC currents [16]-[19] .

In this letter, we propose new multi-cell designs which are able to catch cold atom clouds in a three-dimensional cell-net. Being in the BEC state, the clouds can be entangled due the Josephson effect, and the proposed structure is prospective to demonstrate quantum registering effect regarding to trapped clouds.

---

[1] Extended variant of this letter was submitted to Modern Physics Letters B on February,16 2008

## 2. Nano-Cell Trapping Structures

The most studies in the field of quantum computations are with the realization of entanglement of two sorts of atoms. This is realized by different techniques based on the collisions of atoms in the qubit state [13],[1],[20]. In [21], a new idea was proposed for BECs in a double-well potential. The condensate is governed by the non-linear Schrödinger or Gross-Pitaevsky equation [22], and the Josephson effect arises in the mentioned double-well potential that was confirmed experimentally in [23]. In a system of multi-wells, the qubits can be entangled [24]. Then, a multi-well structure can be a register for a quantum computer[26]. Unfortunately, no such traps have been published yet. In this letter, several nano-cell traps and the potential shapes pertinent for BEC trapping and registering are proposed and studied.

The simplest cell (Fig. 1) consists of two pairs of crossed nanotubes of radius $R_{CN}$ = 3.52 Å. In this figure, the nanotubes are numbered from the left to the right (vertical conductors, $z$-axis), and from bottom to top (horizontal conductors, $y$-axis). Our study of nanotraps orients on the use of the dressing effect for cold alkali atoms. The dressing formalism of the extensive interactions of atoms with an external strong electromagnetic field is proposed in [3]. In this case, atoms interact with a large number of electromagnetic quanta, and several semi-classical approaches were proposed to describe it. In this paper, we study the nanotraps which employ the DC magnetic and RF electromagnetic fields, and the used formulas are based on the theoretical and experimental results [6]. It is taken into account that the cold atoms cannot be placed close to the nonotube surface due to destroying Kasimir-Polder and spin-flip effects, and the cell trapping centers are placed far enough in correspondence with the research results for the same currents and nanotubes.

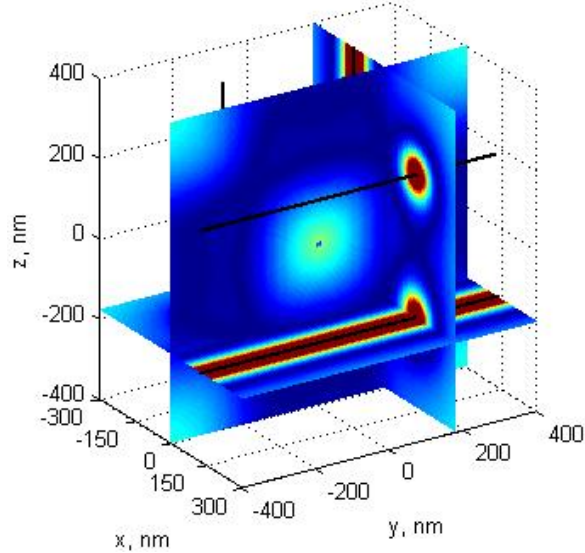

Fig. 1. 3D isolated potential maximum in a four-nanotube cell.

The DC currents in the nanotubes are: $I_{\text{DC}|z-tube}^{(1,2)} = [-15, 15]$ $\mu$A and $I_{\text{DC}|y-tube}^{(1,2)} = [15, -15]$ $\mu$A. The RF currents ($f_{\text{RF}} = 0.27$ MHz) are: $I_{\text{RF}|z-tube}^{(1,2)} = [-4, 4]$ $\mu$A and $I_{\text{RF}|y-tube}^{(1,2)} = [4, -4]$ $\mu$A. The distance between the parallel nanotubes in each layer is $d$ = 355.6 nm, and the layers are placed $h$ = 256.8 nm apart. Minimizing the potential at the four crossing points, it was found that the joint RF and DC excitation of this cell structure provides a local 3D-potential maximum (light-green in the color version) at the center of the cell between the two crossed nanotube pairs (Fig. 1). This maximum is surrounded by a potential minimum (dark-blue) caused by the dressing effect. Directly between the parallel nanotubes on both sides the potential is at an intermediate level which is different from that at the center of the cell. It is presumed that this cell can concentrate the ground state atoms which are seeking the areas with a strong potential. It is also seen that each

nanotube is surrounded by a potential maximum (light-green/red). In the plot, the potential has been limited upwards at $U_{\text{eff}}$ =7e-28 J. This is shown by the brown regions around the nanotubes where the effective potential in reality increases steeply.

The studied trap produces the 3-dimensional effective potential minimum (dark-blue in the color version) at its center if all nanotubes carry only DC currents (Fig. 2). In this case, the potential is calculated as $U_{\text{eff}}(\mathbf{r}) = m_F \mu_B g_F |\mathbf{B}_{\text{DC}}(\mathbf{r})|$. The minimum is completely surrounded by an area with increased potential (light-green). From one side it is glued to the surface of the nanotubes (dark-red).

Thus, this proposed structure can trap two sorts of atoms in the two different regimes described above. The region at the center also appears to stay isolated in all directions while the frequency is reduced in moving from one regime to the other.

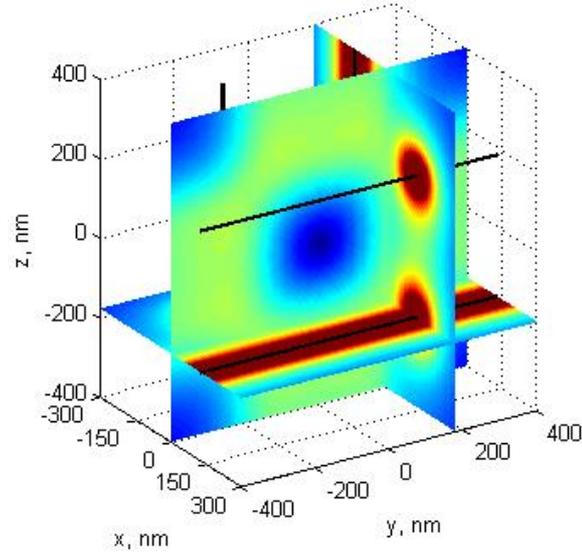

Fig. 2. DC generated effective potential in a four-nanotube cell with the nanotube currents: $I_{\text{DC}}^{(1,2)}\big|_{z-tube} = [-15, 15]$ μA and $I_{\text{DC}}^{(1,2)}\big|_{y-tube} = [15, -15]$ μA.

According to [21] and [24], the qubit effect can arise if the cold atom matter is in the BEC state, and the potential shape consists of two wells, minimum. This theory supposes that a cell structure having several two-well non-overlapping shapes containing the BEC matter can be put into the entanglement state. Our results are on the development of such a cell structure and finding the excitation regime when the multi-extreme potential is realized.

An example of a large structure is shown in Fig. 3 that consists of six pairs of nanotubes of radius $R_{CN}$ = 3.52 Å placed at the distances $h$ = 237 nm, $d_{12}^{(y)} = d_{12}^{(z)} = 337$ nm, $d_{23}^{(y)} = d_{23}^{(z)} = 328.6$ nm and $d_{34}^{(y)} = d_{34}^{(z)} = 329.9$ nm. These six pairs of crossed nanotubes carrying RF and DC currents produce a multi-peak effective potential (Fig. 4).

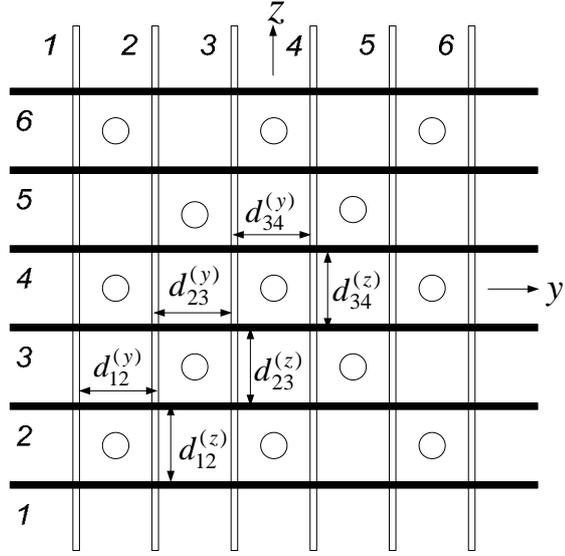

Fig. 3. Cell structure consisting of 6x6 nanotubes. Local potential maxima are indicated with circles.

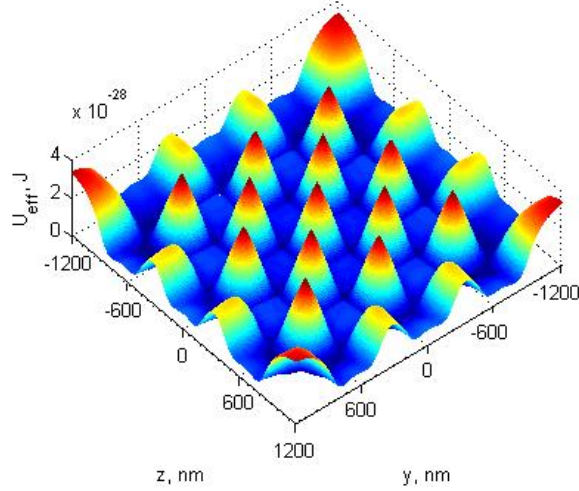

Fig. 4. Effective potential generated by DC and RF currents in the 6x6 nanotube structure. The DC currents in the nanotubes are: $I_{DC|z-tube}^{(1-6)} = [-13.43, 15.08, -15, 15, -15.08, 13.43]$ $\mu$A and $I_{DC|y-tube}^{(1-6)} = [13.43, -15.08, 15, -15, 15.08, -13.43]$ $\mu$A. The RF currents ($f_{RF} = 0.27$ MHz) are: $I_{RF|z-tube}^{(1-6)} = [-4, 4, -4, 4, -4, 4]$ $\mu$A and $I_{RF|y-tube}^{(1-6)} = [4, -4, 4, -4, 4, -4]$ $\mu$A.

The potential is calculated at the plane placed accurately between the two layers of nanotubes. The peaks are isolated three-dimensionally, and it is thought that they can concentrate strong-field-seeking atoms. The cold atoms can be placed at the BEC state, and, according to [21],[24], the matter at the peaks can be entangled under certain conditions.

The same structure produces the multi-minimum potential if the nanotubes carry only DC currents. This is illustrated in Fig. 5 where the 6-pair cell structure is fed by only DC currents of the same values as in Fig. 4. There, the 3D isolated peaks are substituted by isolated potential minima. Additional potential shapes can be derived by further variations of the frequency, DC and RF currents.

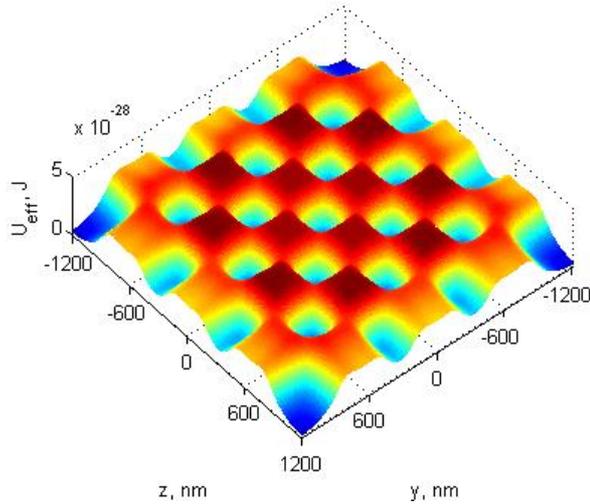
Fig. 5. Effective potential generated by DC currents in the 6x6 nanotube structure.

Crossed nanotube grids such as the one shown in Fig. 4 can be stacked on top of each other. In this case, a more complicated 3D cell-structure is obtained. Our simulations show that the local maxima and surrounding minima become slightly more enhanced toward the center of the structure where the cells are surrounded by other cells. This is illustrated in Fig. 7, where three grids such as in Fig. 4 have been stacked on top of each other. The local potential maxima at the center of the structure are here completely surrounded by a potential minimum. Towards the edges, the local potential maxima are lesser well isolated in the direction between the parallel nanotubes.

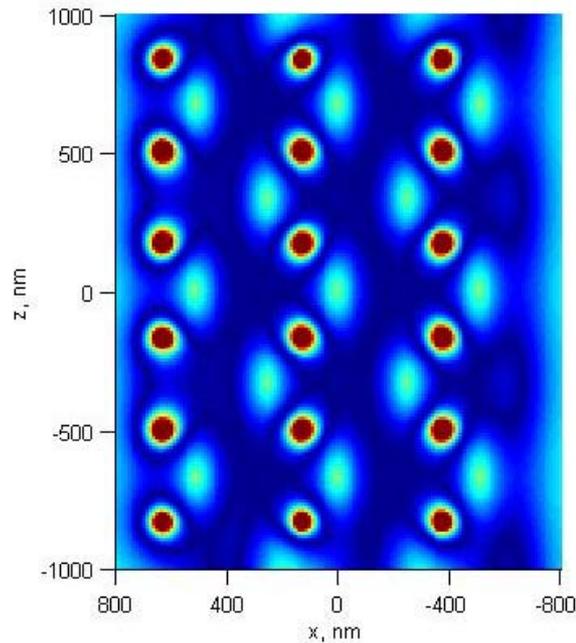
Fig. 6. Slice through a structure where three grids such as in Fig. 3 have been stacked. The shown plane is at the center between nanotube 3 and 4 in the $z$-direction. The nanotubes in the $y$-direction are seen as dark-red circles. The local potential maxima are the light-green ovals between the nanotubes.

Thus, the proposed and studied multi-cell designs show a variety of potential shapes, including the 3D-multi-minima and 3D-multi-maxima depending on the currents in the nanotraps. The steepness of the potential wells can be improved by the use of multi-wall or bundled nanotubes with increased DC and RF currents. The potential barriers between the wells are easily regulated electronically. All results can be scaled towards micron-sized geometries and larger currents.

## 3. Conclusions

In this letter, new three-dimensional multi-cell traps formed by crossed non-touching carbon nanotubes have been developed and studied. The cold dressed atoms are trapped in three-dimensionally isolated cells kept separate by a potential barrier. Tuning of the DC and RF currents allows to place the trapping areas far enough from the domains with strong Casimir-Polder and spin-flip forces, and the traps can confine strong- or low-potential-seeking atoms, depending on the nanotube currents. It has been supposed that these isolated atom clouds being in the BEC state are able to be entangled due to the Josephson effect. The proposed multi-cell structures are additionally interesting in the development of new quantum registers with increased noise immunity and for continuous quantum computations.